\documentstyle[12pt]{article}
\newcommand{\ct}[1]{\cite{#1}}
\newcommand{\np}[1]{Nucl. Phys. {\bf B{#1}}}
\newcommand{\prv}[1]{Phys. Rev. {\bf {#1}}}
\newcommand{\prp}[1]{Phys. Rep. {\bf {#1}}}

\renewcommand{\thefootnote}{\fnsymbol{footnote}}

\newlength{\extraspace}
\newlength{\extraspaces}
\newcommand{\be}{\begin{equation}
\addtolength{\abovedisplayskip}{\extraspaces}
\addtolength{\belowdisplayskip}{\extraspaces}
\addtolength{\abovedisplayshortskip}{\extraspace}
\addtolength{\belowdisplayshortskip}{\extraspace}}
\newcommand{\ee}{\end{equation}}

\newcommand{\ba}{\begin{eqnarray}
\addtolength{\abovedisplayskip}{\extraspaces}
\addtolength{\belowdisplayskip}{\extraspaces}
\addtolength{\abovedisplayshortskip}{\extraspace}
\addtolength{\belowdisplayshortskip}{\extraspace}}
\newcommand{\ea}{\end{eqnarray}}

\newcommand{\nonu}{\nonumber}
\newcommand{\nt}{\noindent}

\newcommand{\la}{\langle}                                 
\newcommand{\ra}{\rangle}                                 

\newcommand{\ie}{{\it i.e., }}

\newcommand{\id}{{\bf 1}}

\newcommand{\bb}{\beta}
\newcommand{\g}{\gamma}

\newcommand{\ep}{\epsilon}

\newcommand{\s}{\sigma}

\newcommand{\gk}{\gamma}

\newcommand{\fk}{\phi}
\newcommand{\Om}{{\C{O}_m}}
\newcommand{\e}{\C{E}}
\newcommand{\C}[1]{{\cal {#1}}}
\newcommand{\B}[1]{\bar {#1}}

\newcommand{\pr}[1]{\partial_{#1}}

\newcommand{\had}[2]{\lim_{{#1} \rightarrow {#2}}~}
\newcommand{\ant}[2]{\int_{\scriptstyle{{#1}}}d^2{#2}~}

\newcommand{\scr}[1]{{\scriptscriptstyle #1}}

\begin{document}
\begin{flushright}
UCLA/93/TEP/48\\
hep-th/9312202\\
December 1993~~~~~~
\end{flushright}

\vskip .5cm

\begin{center}
                   \bf{THE PERTURBATIVE CALCULATION\\
                          OF THE SPIN-SPIN CORRELATION FUNCTION IN\\
                              THE TWO DIMENSIONAL ISING MODEL}
\end{center}
\vskip .75cm

\begin{center}
\bf{Bakhtiar Mikhak}
\footnotesize{and}
\normalsize{\bf{Amir M. Zarkesh\footnote{Supported in part by D.O.E. contract
DE-FG03-91ER 40662 Task C.}}}
\end{center}

\begin{center}
{\it Physics Department\\
University of California, Los Angeles\\
Los Angeles, CA 90024, USA}
\end{center}
\vskip .75cm

\begin{quote}
\hspace{.4cm} Using the variational formula for operator product coefficients
a method for perturbative calculation of the short-distance expansion of the
Spin-Spin correlation function in the two dimensional Ising model is presented.
Results of explicit calculation up to third order agree with known results from
the scaling limit of the lattice calculation.
\end{quote}

\renewcommand{\thefootnote}{\arabic{footnote}}
\setcounter{footnote}{0}
\baselineskip 20pt
\setcounter{equation}{0}
\indent

It has been nearly fifty years since exact solvability and existence of
continuous phase transition in two dimensional Ising model was
shown.\footnote{For the Onsager solution of the Ising model, see ref. \ct{SML}
and references therein.} Ever since then, thanks to equivalence of statistical
mechanics models and quantum euclidean field theories with an ultraviolet (UV)
fixed point, the Ising model has been an excellent theoretical playground for
addressing many field theoretical issues, such as the existence and proper
definition of continuum field theories\ct{WMTB} and the coordinate space
description of the renormalization group (RG). In particular, the existence of
a second order phase transition in the Ising model is equivalent to the
existence of a UV fixed point for the corresponding quantum field theory of a
massive Majorana fermion. Understanding the structure of the short-distance
expansion of correlation functions of the lattice model in the scaling limit by
field theoretical methods amounts to renormalization group analysis and
perturbation theory near the UV fixed point\ct{Wil1}.

In the spirit of classic works of Wilson on the renormalization group and
operator product expansion (OPE) \ct{Wil1}, a new framework for the study of
the theory space of euclidean quantum field theories has been
introduced\ct{Son1}. This formalism is particularly useful for elucidating the
structure of the short-distance expansions of the $n$-point functions of a
renormalizable quantum field theory around a nontrivial fixed point. As a first
application of this formalism a complete renormalization group analysis of the
short-distance expansion of spin-spin correlation function in the scaling limit
of the Ising model has been given\ct{Son2}. The RG analysis determines the
short-distance expansion of the spin-spin correlation function only up to RG
invariant functions. The calculation of these functions requires a perturbative
calculation based on a concrete regularization scheme. The purpose of the
present paper is to show how this calculation is done. To this end we show how
the recursive use of the variational formulas for the correlation
functions\ct{Son4} and operator product coefficients\ct{Son5} leads to a
manifestly finite calculational scheme for the short-distance expansion of the
spin-spin correlation function  in the Ising model.

The scaling limit of the two dimensional massive Ising model is characterized
by the theory of a free massive Majorana fermion which is parametrized by the
cosmological constant and the mass $m$. The cosmological constant only provides
an additional constant free energy density, and therefore none of the
correlation functions of local operators depend on it. Therefore the theory is
parametrized by one real parameter $m$. The mass has scale dimension one and
satisfies the canonical RG equation. Note that $mr$ is RG invariant. The
operator conjugate to $m$, denoted by $\Om$, has the following expectation
value  determined by RG analysis\ct{Son2}

\be \label{Om}
\la \C{O}_m \ra_{m} = -\bb_{\id} m \ln (mc),
\ee

\nt where $\bb_{\id}$ is the anomalous dimension of $\Om$ and $c$ is a constant
of integration. At the fixed point this operator corresponds to the energy
density operator $\C{E}=\frac{1}{2\pi} (i \B{\psi}\psi)$ where $\psi$ is a free
massive Majorana fermion. Our particular choice of normalization fixes
$\bb_{\id}= -\frac{1}{2\pi}$.

The main tool in this scheme is the variational formula which realizes the
derivative of correlation functions with respect to $m$ by the volume integral
of the same correlation functions with an $\Om$ insertion. The necessary
subtractions are made by introducing the operator product expansion from the
outset\ct{Son4}. By counting
{\em scaling\/} dimensions the only necessary subtraction in the present
calculation is

\ba \label{submmid}
A_{mm}{}^\id(\ep;m) &\equiv& - \ant{1 \geq r \geq \ep}{r}
\bar{C}_{mm}^{\scriptscriptstyle{(s)}}{}^\id (r;m) + c_{mm}{}^\id \nonu\\
& =& \frac{1}{2 \pi} \ln (\ep)+{c_{mm}}^\id,
\ea

\nt where $\bar{C}_{mm}^{\scriptscriptstyle{(s)}}{}^\id$ is the singular part
of the coefficient of the OPE of two energy density operators and
$c_{mm}{}^\id$ is the corresponding finite counterterm. Using OPE and the
variational formula for the correlation functions a manifestly finite
variational formula for OPE coefficients can be derived\ct{Son5}. We show how
this VF leads to a manifestly finite scheme for the perturbative calculation of
the corrections to the scaling behavior of the spin-spin correlation function.

Consider the short distance expansion for spin-spin correlation

\be \label{Cssk}
\la \s (r) \s (0) \ra_m =C_{\s \s}{}^b (r;m) \la \fk_b \ra_m,
\ee

\nt where the sum over $b$ is sufficient to include scalar composite operators,
which are not total derivatives.  Because of the operator algebra of the
critical Ising model, $\fk_b \in \{[\id],[\e]\}$. To calculate the corrections
to scaling behavior of the OPE coefficients to third order\footnote{Up to scale
dimension 3 the only non-derivative scalar operators are $\id~$ and
$\C{E}$\ct{Son2,BPZ}.} in mass it is sufficient to truncate the sum over $b$
to\footnote{For a general proof of truncation of the variational formula to
arbitrary order see ref. \ct{MZ1}.}

\be \label{Csskt}
\la \s (r) \s (0) \ra_m =C_{\s \s}{}^\id (r;m) + C_{\s \s}{}^m (r;m) \la \Om
\ra_m.
\ee

The zero-th order is just the well-known scaling limit. In first order the
following {\em finite\/} and {\em truncated\/} first order VF for the OPE
coefficients can be derived\ct{Son5,MZ1}

\ba \label{OPEex1}
\lefteqn{-\pr{m}C_{\s\s}{}^{\id}(r)= \had{R'}{\infty} \bigg{\{}
\ant{\scr{R'\geq r'}}{r'}  \la\e(r')\s(r)\s(0)\ra} \nonu \\
&~~~~~~~~~~~~~~~&~~~~~~~~~~- C_{\s\s}{}^m(r)~ \had{\ep}{0} \Big(
\ant{\scr{R'\geq r'\geq\ep}}{r'}  \la\e(r')\e(0)\ra +A_{mm}{}^{\id}(\ep) \Big)
\bigg{\}}.~~~~~
\ea

The free massive Majorana fermion representation for the critical Ising
model\ct{BPZ} makes the calculation of the energy-density correlation functions
trivial. The expression of $\s(r)$ in terms of the fermion field, on the other
hand, is non-local. This makes the calculation of the correlation functions,
which involve the spin operator, {\em non-trivial}\footnote{This is the main
reason that, even though the critical Ising model has a representation in terms
of a free massless Majonara fermion, it is still a {\em non-trivial\/} fixed
point theory.}. In two dimensions, all correlation functions of conformally
invariant field theories are calculable in principle\ct{BPZ}. Moreover, in the
case of the critical Ising model there are additional structures which give a
systematic way of {\em recursively} calculating the correlation
functions\ct{DSZ} needed in the main terms of the VFs (\ie the terms denoted by
$\C{M}$s in what follows). In the following we use the results, which, using
the techniques of \ct{DSZ}, can be easily derived\ct{MZ1}. For example, in the
case of VF (\ref{OPEex1}) we have

\ba \label{OPEexx1}
-\pr{m}C_{\s\s}{}^{\id}(r)&=& \had{R'}{\infty} \bigg{\{}
r^{\frac{3}{4}}~\C{M}_1\Big(\frac{R'}{r}\Big)- C_{\s\s}{}^m(r) \bigg(
\C{S}_{1,\e}\Big(\frac{R'}{r}\Big) +c_{mm}{}^\id \bigg) \bigg{\}}.
\ea

\nt where

\be
\C{M}_1(U')=\int_{\scr{U' \geq u'}} \frac{d^2u'}{4 \pi |u'||u'-1|},
\ee
\be
\C{S}_{1,\e}(U') = \had{\ep}{0} \Big( \int_{\scr{U'\geq u'\geq \frac{\ep}{r} }}
\frac{d^2u'}{4 \pi^2 |u'|^2} +\frac{1}{2 \pi} \ln (\ep) \Big).
\ee

The repeated use of variational formula leads to expressions for higher order
derivatives of $C_{\s\s}{}^\id$ evaluated at $m=0$. In deriving these
expressions up to third order we should keep the following two points in mind.
First, since we only consider VF for correlation functions which involve
energy-density and spin operators, the knowledge of the subtraction
(\ref{submmid}) is sufficient. Second, in second and third order VFs, where the
arguments of two inserted energy-density operators, one coming from a lower
order VF, approach one another, we do not need any subtractions at all. This is
due to the {\em connectedness\/} of the insertions, \ie we have to consider
{\em partially\/} connected correlation functions with respect to {\em all\/}
insertions. Therefore, by taking the derivative of VF (\ref{OPEex1}) and using
the first order VF for correlation functions, we could derive the following
second order finite and truncated VF

\ba \label{OPEt2}
\lefteqn{\pr{m}^2 C_{\s\s}{}^{\id}(r)= \had{R'}{\infty} \had{R''}{\infty}
\bigg{\{ } \ant{\scr{R'\geq r'}}{r'} \ant{\scr{R''\geq r''}}{r''} \la \e(r'')
\e(r') \s(r) \s(0) \ra^c} \nonu \\
& &+ \pr{m} C_{\s\s}{}^m(r) \bigg[ \had{\ep}{0} \Big( \ant{\scr{R'\geq
r'\geq\ep}}{r'} \la \e(r') \e(0) \ra +A_{mm}{}^{\id}(\ep) \Big) + (R'
\leftrightarrow R'') \bigg] \bigg{\} },~~
\ea

\nt The effect of the terms corresponding to $\pr{m}^2 \la \e \ra$ vanish
because of the specific order of the limits and the OPE algebra of the critical
Ising model. We emphasize that this equation is finite\footnote{For more
detailed explanation of this subtle point see ref. \ct{MZ1}.}. Using the
explicit expression for the necessary four point function we have

\ba \label{OPEt2e}
\lefteqn{\pr{m}^2 C_{\s\s}{}^{\id}(r) = \had{R'}{\infty} \had{R''}{\infty}
\bigg{\{ } } \nonu \\
& & ~~~~~~~~~~~r^{\frac{7}{4}} ~\C{M}_2 \Big(\frac{R'}{r}, \frac{R''}{r}\Big)+
\pr{m} C_{\s\s}{}^m(r) \bigg[ \C{S}_{1,\e}\Big(\frac{R'}{r}\Big) +
\C{S}_{1,\e}\Big(\frac{R''}{r}\Big) + 2 {c_{mm}}^\id \bigg] \bigg{\} },
\ea

\nt where

\ba
\lefteqn{\C{M}_2 (U',U'') = \int_{\scr{U'\geq u'}} \frac{d^2u'}{4 \pi
|u'||u'-1|} \int_{\scr{U''\geq u''}}\frac{d^2u''}{4 \pi |u''||u''-1|} } \nonu\\
& & ~~~~~~~~~~~~~~~~~~ \times \bigg[
\frac{|u'+u''-2u'u''|^2-4|u'||u'-1||u''||u''-1|}{|u'-u''|^2} \bigg].~~~~~
\ea

\nt In complete analogy to second order we can derive higher order finite and
truncated variational formulas\ct{MZ1}. All the necessary critical correlation
functions can be calculated with the aid of the recursive relations of
ref.\ct{DSZ}. As a result we have two types of integrals to handle. The
calculation of the subtraction integrals $\C{S}$'s are straightforward, due to
the rationality of the integrands. But, the main integrals $\C{M}$'s are in
general quite hard and have been one of the main obstacles for the field
theoretical calculation of the short-distance expansion of the spin-spin
correlation function. Here we give a general method for calculating these
integrals\ct{MZ1}.

Taking the IR limits to infinity has a very important calculational advantage;
usually it is much easier to calculate integrals for infinitely large cutoffs
rather than finite cutoffs. Taking advantage of this fact the integral
$\C{M}_1$ could be calculated and involves elliptic functions. But, the higher
order integrals are very difficult to calculate directly. The main reason for
this difficulty is the appearance of one copy of the integrand of $\C{M}_1$,
for each integration. Once we analyze the structure of $\C{M}_1$ carefully, the
analysis of the higher order integrals follows.

The separation of the two poles of the integrand of $\C{M}_1$ makes the
integral asymmetric respect to the IR and the UV behavior; we do not have any
UV divergences in the main term. Therefore we have to look for a transformation
which symmetrizes the integral with respect to the IR and the UV limits. We now
demonstrate that the following conformal transformation does the job

\be \label{sym}
u'=\frac{1}{2} \Big( 1+\frac{1}{2}(x+\frac{1}{x}) \Big);
\ee

\nt hence, the name {\em symmetrization} transformation. Under this
transformation, for large cutoff, the $\C{M}_1$ becomes

\be
\C{M}_1(U') = \int_{\scr{4U' \geq x \geq 1}} \frac{d^2x}{4 \pi |x|^2}~ +
O(\frac{1}{U'}).
\ee

After applying the symmetrization transformation, the integral in
(\ref{OPEexx1}) is trivial and for $c_{mm}{}^\id = \frac{1}{\pi}\ln2$ we get
the following results

\ba \label{aha}
C_{\s\s}{}^m(r)&=& \pi r^{\frac{3}{4}}, \\
\label{ahaa}
\pr{m}C_{\s\s}{}^\id (r)&=& r^{-\frac{3}{4}} \ln r.
\ea

\nt The logaritmic dependence on $r$ in (\ref{ahaa}) could be also derived from
solving the RG equation\ct{MZ1}. In addition, using the VF for $\la \Om \ra_m$,
and the representation of Ising model in terms the theory of the free massive
fermions it is possible to calculate the complete relation between $c$ in
(\ref{Om}) and $c_{mm}{}^\id $\ct{MZ1}. The final answer for our choice of
${c_{mm}}^{\id}$ is $c = \frac{e^\g}{8}$, where $\gk$ is the Euler constant.
This completes the specification of our choice of the conjugate operator. The
first non-trivial information will come out of the second order calculation.

The calculation of $\C{M}_2$ involves non-trivial integration of elliptic
functions and is extremely hard.\footnote{Dotsenko applied his conformal field
techniques to calculate this integral by clever contour deformations and
analytic continuation methods.\ct{Dot} But, as he also points out, generalizing
his technique to higher orders is very difficult.}. But using the
symmetrization transformation, in the limit of large cutoffs, this integral is
calculable with standard methods and we get

\ba \label{sadeh}
\had{U'}{\infty} \had{U''}{\infty} \C{M}_2 (U',U'') = \frac{1}{8}.
\ea

\nt Therefore, the main term in (\ref{OPEt2e}) does not have any IR
divergences, and we have

\ba \label{zero}
\pr{m} C_{\s\s}{}^m(r)&=&0 ~, \\ \label{eight}
\pr{m}^2 C_{\s\s}{}^\id (r)&=& \frac{r^{\frac{7}{4}}}{8}.
\ea

\nt Note that, (\ref{zero}) is a result independent of the choice of any finite
counterterms. In ref.\ct{Son2} this was a result expected from comparison to
the exact result of ref.\ct{WMTB}.

The third order calculation also relies on the symmetrization transformation.
Just like the second order calculation, the result of applying this
transformation is an integral which can be handled by standard techniques. We
finally get

\ba \label{sixteen}
\pr{m}^2 C_{\s\s}{}^m (r)&=& \frac{\pi r^{\frac{11}{4}}}{8},\\ \label{third}
\pr{m}^3 C_{\s\s}{}^\id (r)&=& \frac{3r^{\frac{11}{4}}}{16} \ln r.
\ea

\nt Contrary to (\ref{zero}), the vanishing result for the constant part of
(\ref{third}) is due to the specific choice of the finite counterterm
$c_{mm}{}^\id$. We can now calculate the short-distance expansion on the
spin-spin correlation function to third order in the mass. In complete
agreement with the scaling limit of the exact lattice results of ref.\ct{WMTB}
we have

\be \label{2spin}
\la \s(r)\s(0) \ra_m = \frac{1}{r^{\frac{1}{4}}} \Big( 1 + \frac{1}{2} t \ln
(|t|e^{\g}/8) + \frac{1}{16} t^2 + \frac{1}{32} t^3 \ln (|t|e^{\g}/8) +
O(t^4\ln ^2|t|) \Big),
\ee

\nt where $t \equiv mr$. This is the first {\em systematic\/} field-theoretic
calculation near a {\em non-trivial\/} fixed point.

Note that our method is general, and in principle, there seems to be no
difficulty in continuing the calculation to higher orders. A complete proof of
the truncation, as well as a systematic study of the structure of the higher
order calculations, applicability of the symmetrization transformation to
arbitrary order, and details of the calculations, is presented
elsewhere\ct{MZ1}. We believe that our methods should also be applicable for
perturbative calculations around other non-trivial fixed points corresponding
to more complicated minimal models classified in ref.\ct{BPZ}. It is important
to note that the proof of the finiteness and the possibility of truncation is
applicable to all models of this type\ct{MZ1}. Clearly, there is much important
work to be done in this direction.

We would like to thank H. Sonoda for suggesting the problem, his valuable
comments, and constant encouragement.



\baselineskip 10pt
\begin{thebibliography}{99}
%
\bibitem{SML}
T.D. ~Schultz, D.C. ~Mattis, and E.H. ~Lieb, Rev. Mod. Phys. {\bf 36}(1964)
856. %
\bibitem{WMTB}
T.T. ~Wu, B.M. ~McCoy, C.A. ~Tracy, and E. ~Barouch, Phys. Rev. {\bf B13}(1976)
316.
%
\bibitem{Wil1}
K. ~Wilson, \prv{179}(1969) 1499.\\
K. ~Wilson and J. Kogut, \prp{12C}(1974) 76.
%
\bibitem{Son1}
H. ~Sonoda, \np{352}(1991) 585.
%
\bibitem{Son2}
H. ~Sonoda, \np{352}(1991) 601.
%
\bibitem{Son4}
H. ~Sonoda, \np{383}(1992) 173.
%
\bibitem{Son5}
H. ~Sonoda, \np{394}(1993) 302.
%
\bibitem{MZ1}
B. ~Mikhak and A.M. ~Zarkesh, {\it Manifestly Finite Perturbation Theory for
the Short-distance Expansion of Correlation Functions in the Two Dimensional
Ising Model}
Prepint UCLA/93/TEP/49
%
\bibitem{Dot}
V.I. ~Dotsenko, \np{314}(1989) 687.
%
\bibitem{BPZ}
A.A. ~Belavin, A.M. ~Polyakov, and A.B. ~Zamolodchikov, \np{241}(1984) 333.
%
\bibitem{Weg}
F. ~Wegner, Phys. Rev. {\bf B5}(1972) 4529; J. Phys. {\bf A8}(1975) 710.
%
\bibitem{DSZ}
P. ~Di Francesco, H. ~Saleur, and J.B. ~Zuber \np{290}(1987) 527.
%
\end{thebibliography}
\end{document}